\documentclass[aps,floats,showpacs]{revtex4}

\usepackage[dvips]{color}     %serve per i colori
\usepackage{epsfig,graphicx}
\newcommand{\be}{\begin{equation}}
\newcommand{\ee}{\end{equation}}
\newcommand{\bea}{\begin{eqnarray}}
\newcommand{\eea}{\end{eqnarray}}
\newcommand{\beas}{\begin{eqnarray*}}
\newcommand{\eeas}{\end{eqnarray*}}
\newcommand{\bt}{\begin{tabular}}
\newcommand{\et}{\end{tabular}}
\newcommand{\ba}{\begin{array}}
\newcommand{\ea}{\end{array}}

\newcommand{\noi}{\noindent}
\newcommand{\drm}{{\rm d}}
\newcommand{\erm}{{\rm e}}

\begin{document}

%---------------------------------------------------------------------------

\title{A new dielectric effect in viscous liquids} %{\em Pipe effect}

\author{V. Capano}
\affiliation{Universit\`a Statale di Bergamo, Facolt\`a di
Ingegneria, viale Marconi 5, I-24044 Dalmine (BG), Italy }
\author{S. Esposito}
\email{Salvatore.Esposito@na.infn.it}%
\affiliation{Istituto Nazionale di Fisica
Nucleare, Sezione di Napoli, Complesso Universitario di Monte S.
Angelo, via Cinthia, I-80126 Naples, Italy}
\author{G. Salesi}
\email{Giovanni.Salesi@unibg.it} %
\affiliation{Universit\`a Statale di Bergamo, Facolt\`a di
Ingegneria, viale Marconi 5, I-24044 Dalmine (BG), Italy \\ and
Istituto Nazionale di Fisica Nucleare, Sezione di Milano, via
Celoria 16, I-20133 Milan, Italy}

\begin{abstract}
\noindent An accurate experimental and theoretical study has been
performed about a phenomenon, not previously reported in the
literature, occurring in highly viscous liquids: the formation of
a definite pipe structure induced by the passage of a heavy body,
this structure lasting for quite a long time. A very rich
phenomenology (including mechanical, optical and structural
effects) associated with the formation of the pipe has been
observed in different liquids. Actually, the peculiar dynamical
evolution of that structure does not appear as a trivial
manifestation of standard relaxation or spurious effects. In
particular we have revealed different time scales during the
evolution of the pipe and a non-monotonous decrease of the
persistence time with decreasing viscosity (with the appearance of
at least two different maxima). We put forward a microscopic model, consistent
with the experimental data, where the pipe behaves as a
``dielectric shell'' whose time evolution is described through a 
simple thermodynamical approach, predicting several properties
effectively observed.

%\pacs{68.18.Fg; 68.05.Cf; 61.20.Lc; 77.22.Gm; 77.55.+f; 77.22.-d}
%47.55.dr Interactions with surfaces
%47.27.nf Flows in pipes and nozzles
%51.70.+f Optical and dielectric properties
%77.22.-d Dielectric properties of solids and liquids
%77.22.Gm Dielectric loss and relaxation
%61.20.Lc Time-dependent properties; relaxation
%66.20.-d Viscosity of liquids; diffusive momentum transport
%68.03.Hj Liquid surface structure: measurements and simulations
%68.05.-n Liquid-liquid interfaces
%68.05.Cf Liquid-liquid interface structure: measurements and simulations
%68.18.Fg Liquid thin film structure: measurements and simulations
\end{abstract}

%77.22.-d Dielectric properties of solids and liquids
%36.40.-c Atomic and molecular clusters 
%64.60.My Metastable phases  
%64.70.Ja Liquid-liquid transitions  
%64.70.pm Liquids  

\maketitle

%\vskip2pc]

%---------------------------------------------------------------------------

\section{Introduction}

\noindent One of the most active area of research in soft
condensed matter physics, physical chemistry, materials science
and biophysics is the study of dynamics of complex systems and of 
its relationship to their structure. Among such complex systems, a
special place is occupied by hydrogen-bonding liquids and their
mixtures \cite{HBond1} \cite{HBond2}, because of their extreme
prominence in different biological and technological processes.
This is especially true for glycerol (C$_3$H$_8$O$_3$), where the
presence of three hydroxyl groups per molecule makes it a
particularly rich and complex system for the study of hydrogen
bonded fluids. Glycerol, indeed, has been the subject of
considerable and long-standing scientific interest \cite{Glyc12}
due to its complex nature.
For example, the peculiarities of the intermolecular interaction
of glycerol with water via hydrogen bonds form the basis of the
valuable hydration properties of glycerol, that are widely applied
in the pharmaceutical, cosmetics and food industries. Just to
quote typical examples, glycerol has been employed as a
cryoprotector \cite{cryopro}, depending on the changes
of the parameters of the phase transitions of water in the
presence of glycerol, but it has been also the focus of study by
researchers in cryopreservation \cite{cryopre}.
On the other hand, glycerol is an excellent glass former, and has
been extensively studied experimentally \cite{glassformer} in
connection with attempts to understand the nature of the glass
transition. Understanding the liquid-glass transition and its
related dynamics, in fact, is one of the most important and
challenging problems in modern condensed matter physics, and
glycerol and its mixtures with water are widely used as models to
study the cooperative dynamics, glass transition phenomena and
scaling properties in complex liquids.

The dielectric properties of hydrogen-bonded liquids are, as well,
of key interest, because such liquids generally show anomalous
dielectric behavior, which is not observed in a
non-hydrogen-bonded liquid. In fact, with some exception (such as
acetic acid), the static dielectric constant of a hydrogen-bonded
liquid is generally larger than that of a normal polar one, mainly
because of the regular alignment of a dipolar molecule in the
hydrogen-bonded cluster. This applies especially to glycerol,
where the existence of long- and short-range forces, and a large
variety of possible molecular conformations, leads to important
dynamics on a variety of time scales.
All these facts evidently motivate, on one hand, the large
research effort about hydrogen-bonded liquids and, in particular,
glycerol, but also urge to study further such systems, searching
for possible novel phenomena that put some other light on their
interesting physical properties. In the present paper we
extensively report just on an apparently new effect (which in the
following we shall often call as ``pipe-effect'') of such a kind
observed in glycerol that, seemingly, has not been considered
before (at least in the published literature).

The starting observation is strictly related to the standard
determination of the viscosity by means of the method of the
falling sphere. Although not easily seen with the naked eye, after
the falling of the metal sphere in glycerol at common
temperatures, a pipe appears in the viscous liquid that persists
for some long time. At a first glance, such a not surprising
effect seems to be easily explained in terms of some relaxation
processes occurring in highly viscous media. However, the lacking
of an apparent microscopic structure in liquids, though highly
viscous, similar to solids, seems mainly to disfavor the simple
explanation envisaged above, while requiring more observations on
the phenomenon. Indeed, such observations do reveal a very rich
phenomenology, pointing out that the effect observed is, quite
surprisingly, not at all trivial.

\begin{figure}%1 NUOVA FIGURA 1
\begin{center}
\epsfxsize=3.5cm %
\epsffile{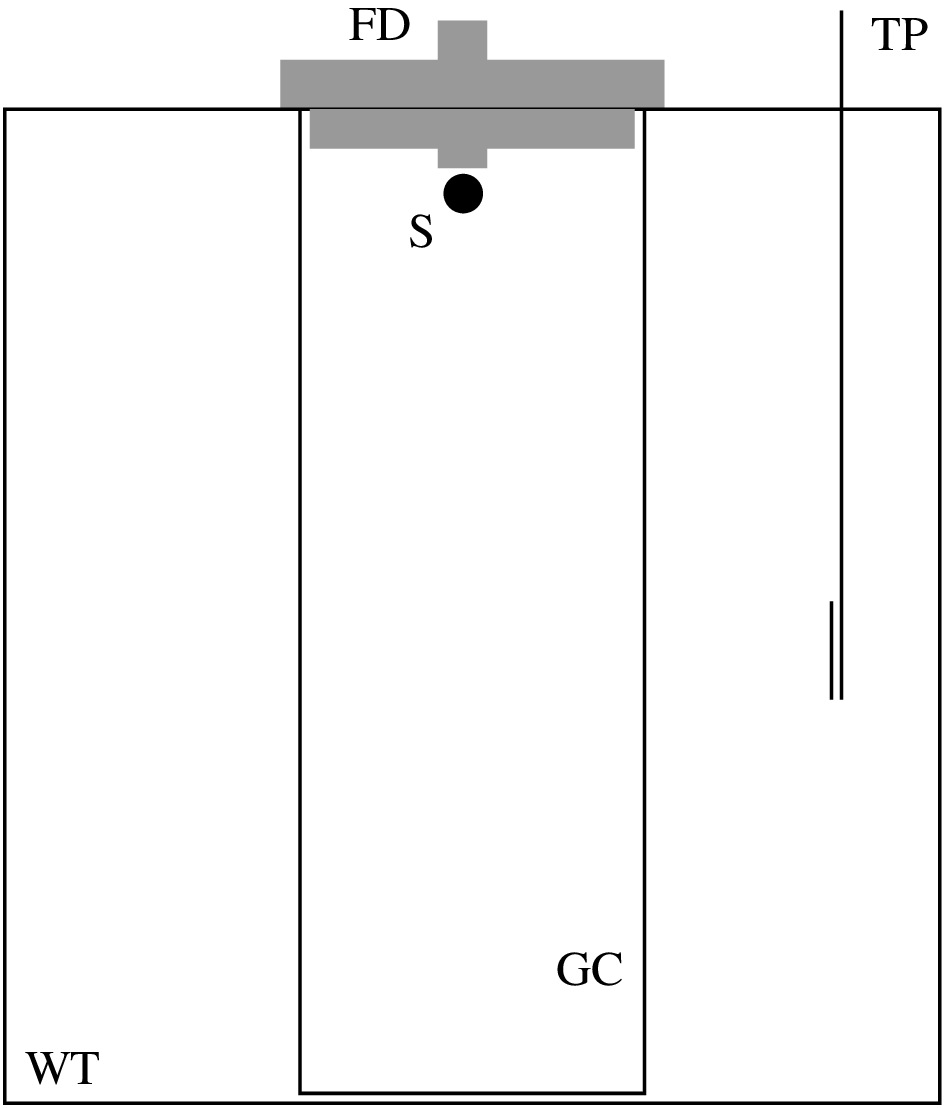} a) %\\ ${}$   %glycsetup.eps
\qquad\quad
\epsfxsize=8cm %
\epsffile{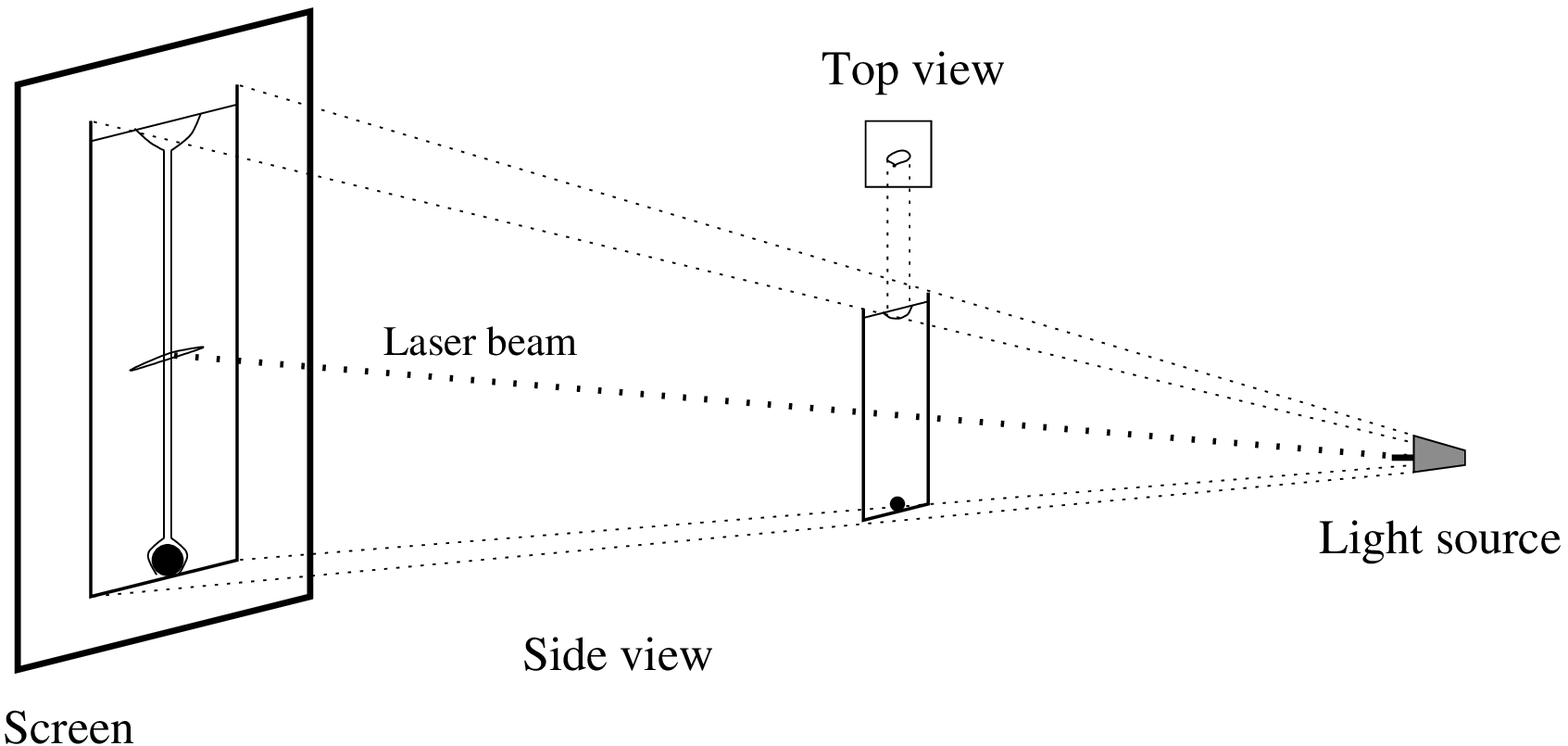} b)            %glycprojec.eps
\caption{a) {\em experimental apparatus employed to produce and observe the pipe effect in glycerol: {\rm WT}, water thermostat; {\rm GC}, glycerol container (graduated tube); {\rm FD}, falling device; {\rm S}, steel sphere; {\rm TP}, temperature probe}; \ b) {\em a schematic view of the experimental arrangement for studying the optical properties of the pipes formed in glycerol.}}
\label{Fig1}  
%\label{glycsetup}=>fig.1a
%\label{glycprojec}=>fig.1b
\end{center}
%\vspace*{-1.7truecm}
\end{figure}

We have thus started a set of appropriate experiments aimed at
collecting all the relevant phenomenology (both qualitative and
quantitative), upon which a theoretical explanation of the
phenomenon, though preliminary, can be consistently built. The
results of our preliminary study are reported here. After a
description in Section II of all the experimental observations
obtained, in Section III we give a
theoretical interpretation of the phenomena observed. In
particular, we propose a possible microscopic interpretation of
the occurrence of the pipe effect, discussing in detail the
corresponding theoretical model, and propose and solve a thorough
thermodynamical model which describes the evolution of the pipe, 
while deducing some of the properties observed. Finally, in 
Section IV, we summarize the results obtained and give our 
conclusions and outlook.

\section{Experimental observation}

\noindent In the experiments performed we have employed different
commercial viscous liquids, glycerol (C$_3$H$_8$O$_3 \geq 99.5
\%$, water content $\leq 0.1 \%$), ethanol (C$_2$H$_6$O $\geq 96
\%$) and castor oil (essentially pure), as delivered, while, for
the mixtures with water, a doubly distilled water was used. We
have focused our attention particularly to glycerol. For the sake 
of simplicity, in the following we will refer to ``pure'' glycerol 
when no solute (water or ethanol) has been added; we have checked 
that, in such case, the concentration $x$ of glycerol is about 
$x=0.99$ (mainly due to pre-existing impurities), that will then 
be our reference value.

The several experiments carried out consisted, basically, always
in inducing the formation of the abovesaid pipe in the 
viscous liquid, just by means of the falling (from the rest) of 
steel spheres of different diameters, and then observing various 
properties about the formation and evolution of the pipe, or 
measuring different properties of the pipe itself. A sketch of 
the experimental setup employed is roughly depicted in 
Fig.\,\ref{Fig1}a. For obvious reasons, we have taken some care 
in keeping constant and uniform the relevant parameters (pressure, 
temperature, concentration, etc.) of the probing liquids, contained 
in given graduated tubes (heights ranging from 15 to 30 cm, diameters 
from 2 to 5 cm).

While the formation of the pipe follows closely the falling of the
sphere (it, is, practically, instantaneous), its disappearance
after some time is not, in the given experimental conditions,
unambiguous. In order to keep the errors on the persistence time
of the phenomenon under control, we have always adopted the same
protocol to mark the ``complete'' disappearance of the pipe (that
is, when the luminosity of the pipe with respect to the bulk
liquid falls below a given value, necessarily different from
zero).
For the study of the optical properties of the pipe, we have employed an arrangement schematically depicted in Fig.\,\ref{Fig1}b. Light from a standard filament, or from halogen lamps, or from a common He-Ne laser, impinges on the pipe, and the image formed is collected on an opaque screen far away from the graduated tube containing the viscous liquid, thus obtaining a lateral view of the pipe itself. Instead, a top view of the pipe (which is obtained even without specific illumination from the light source) is simply obtained by looking from the top of the container.

The properties of the pipe have been studied in four different
kinds of viscous liquids, namely: pure glycerol, glycerol/water
mixtures and glycerol/ethanol mixtures with different
concentrations, and pure castor oil. The motivation for this
choices lies in their structural properties and, in particular, on
the values of two main quantities of those substances, that are
viscosity and surface tension. Indeed, glycerol has very high
viscosity ($\eta_{\rm glycerol} = 1410 \times 10^{-3}$ Pa $\!\cdot\!$ s at
$20^{\mathrm o}$C) and surface tension
($\sigma_{\rm glycerol} = 63.4 \times 10^{-3}$ N/m at $20^{\mathrm
o}$C), while water and ethanol have a low viscosity ($\eta_{\rm water}
= 1.002 \times 10^{-3}$ Pa $\!\cdot\!$ s and $\eta_{\rm ethanol} = 1.200
\times 10^{-3}$ Pa $\!\cdot\!$ s at $20^{\mathrm o}$C) but high
($\sigma_{\rm water} = 72.8 \times 10^{-3}$ N/m at $20^{\mathrm o}$C)
and intermediate ($\sigma_{\rm ethanol} = 22.8 \times 10^{-3}$ N/m at
$20^{\mathrm o}$C) surface tension, respectively, so that, with
different mixtures, at least the dependence on viscosity and
surface tension of pipe formation and evolution can be
discriminated. The use, in further experiments, of castor oil too,
which have an high viscosity, comparable to that of pure glycerol
($\eta_{\rm castor \, oil} = 986 \times 10^{-3}$ Pa $\!\cdot\!$ s), but a completely different molecular structure, may shed some light on the influence of the structure on the phenomenon.

\

\noi Just after the falling of spheres of different diameters in pure glycerol, irrespective of the starting falling conditions (from outside the tube or from inside the glycerol), the diameters (from 2 mm to 12 mm) and the shape (cylinder or prism with different basis) of the tubes, a pipe is formed in the viscous liquid. The same applies to glycerol/water mixtures (with concentration as low as
$x_{\rm glycerol} = 0.80$) and to glycerol/ethanol
mixtures \footnote{Note that even in glycerol/ethanol mixtures a small content of spurious water is present, not lower than $1\%$,
mainly due to to the content of water in the ``pure'' glycerol
employed.} (with concentration as low as $x_{\rm glycerol} = 0.90$), as well as to castor oil, although in a less marked way. The main general features of the pipe formed in the different liquids employed are similar, that is a funnel-shaped upper part sharply narrowing to a cylinder towards the bottom. In glycerol/water and glycerol/ethanol mixtures, the shape of the pipe is not, however, regularly cylindrical (with a definite straight line axis) as in the case of pure glycerol, but, nevertheless, for glycerol/ethanol mixtures (for any concentration considered) a more pronounced cylindrical structure appears (just as a ``tube immersed'' in the bulk liquid). In some cases, when employing falling spheres with large diameters (around
10 mm), a double pipe structure is even observed: a pipe with lower
diameter is inside a larger pipe. \footnote{Illuminated by normal (not laser) light, the smaller pipe appears on a screen with a darker shadow with respect to that of the larger
one, though always darker with respect to the bulk glycerol.
Quite rapidly, however, the outer pipe gets narrower till it
coincides with the inner one.}

The pipe effect is observed (in a very visible way) even if, instead of falling steel spheres, water droplets on the bottom of the tube are used in their re-climbing motion in pure glycerol. In such a case, however, the pipe deforms very rapidly (since water is soluble in glycerol) and, after a given time, all the water is absorbed by glycerol, leaving no track of its passage. The pipe, instead, does not form if a (macroscopic) air bubble is used during its re-climbing motion in pure glycerol.
We have also noted that the phenomenon is not at all observed (in any kind of tubes, in vertical or inclined positions) if, before falling, the spheres are deposited in the viscous liquid for a very long time (greater than five hours).

Microscopic air bubbles are always present in the pipe along its full length when the steel sphere is left to fall in the viscous liquid starting from a point outside it (i.e., in air), this being obviously due to the adsorption of air by the sphere. Such micro-bubbles are practically trapped inside the pipe (they climb up very slow), and render much more visible the pipe when illuminated.
However, such secondary effect (as well as others mentioned below) may easily hide the primary phenomenon concerning the ``true'' dynamics of the pipe, so that we have taken particularly care of avoiding such spurious effects (whenever possible) in performing our experimental measurements.

\begin{figure}[!t]%5 NUOVA FIGURA 2
\begin{center}
\begin{tabular}{ccc}
\epsfxsize=4.3cm %
\epsffile{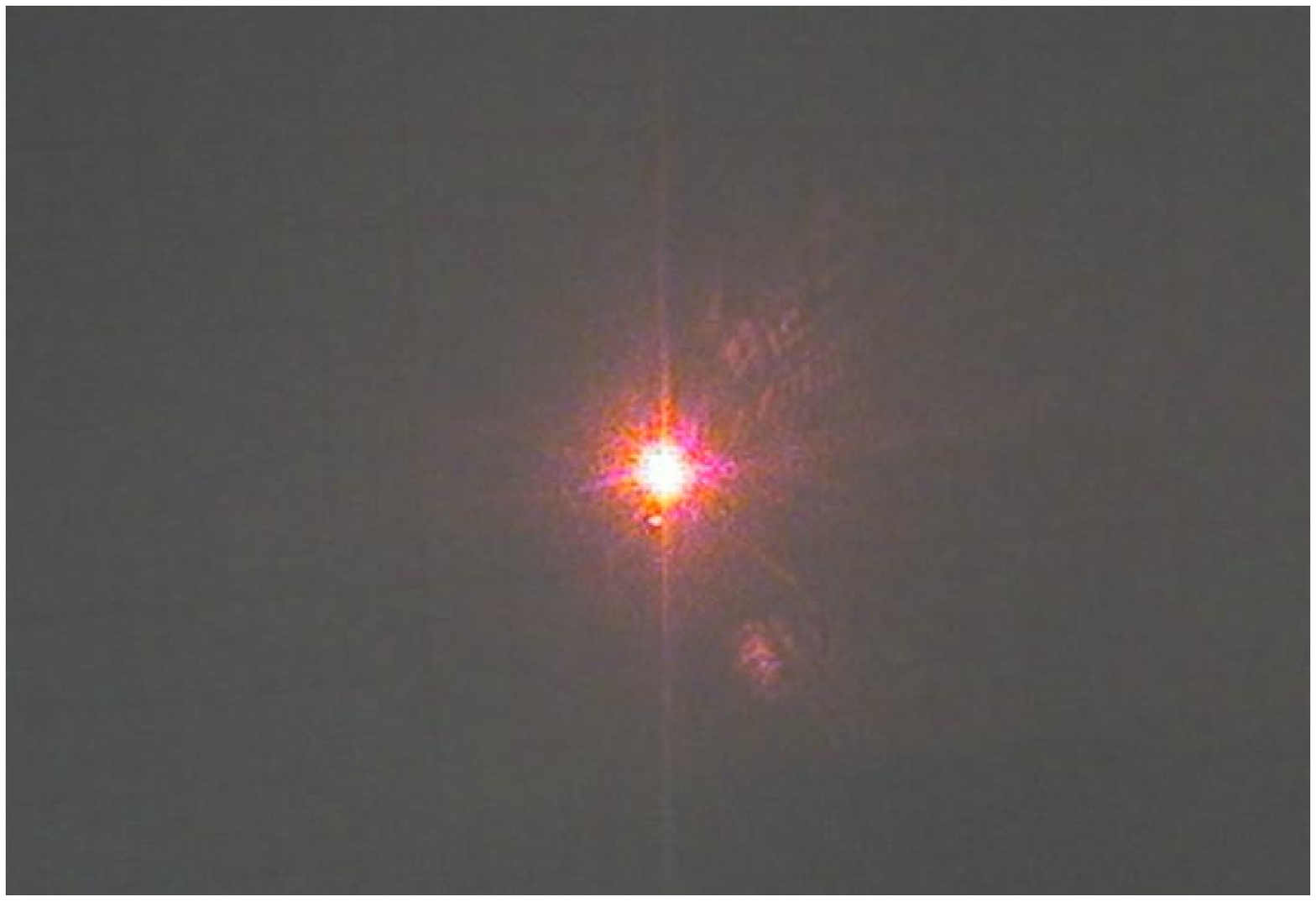} a) & & %bulk.eps
\epsfxsize=4.3cm %
\epsffile{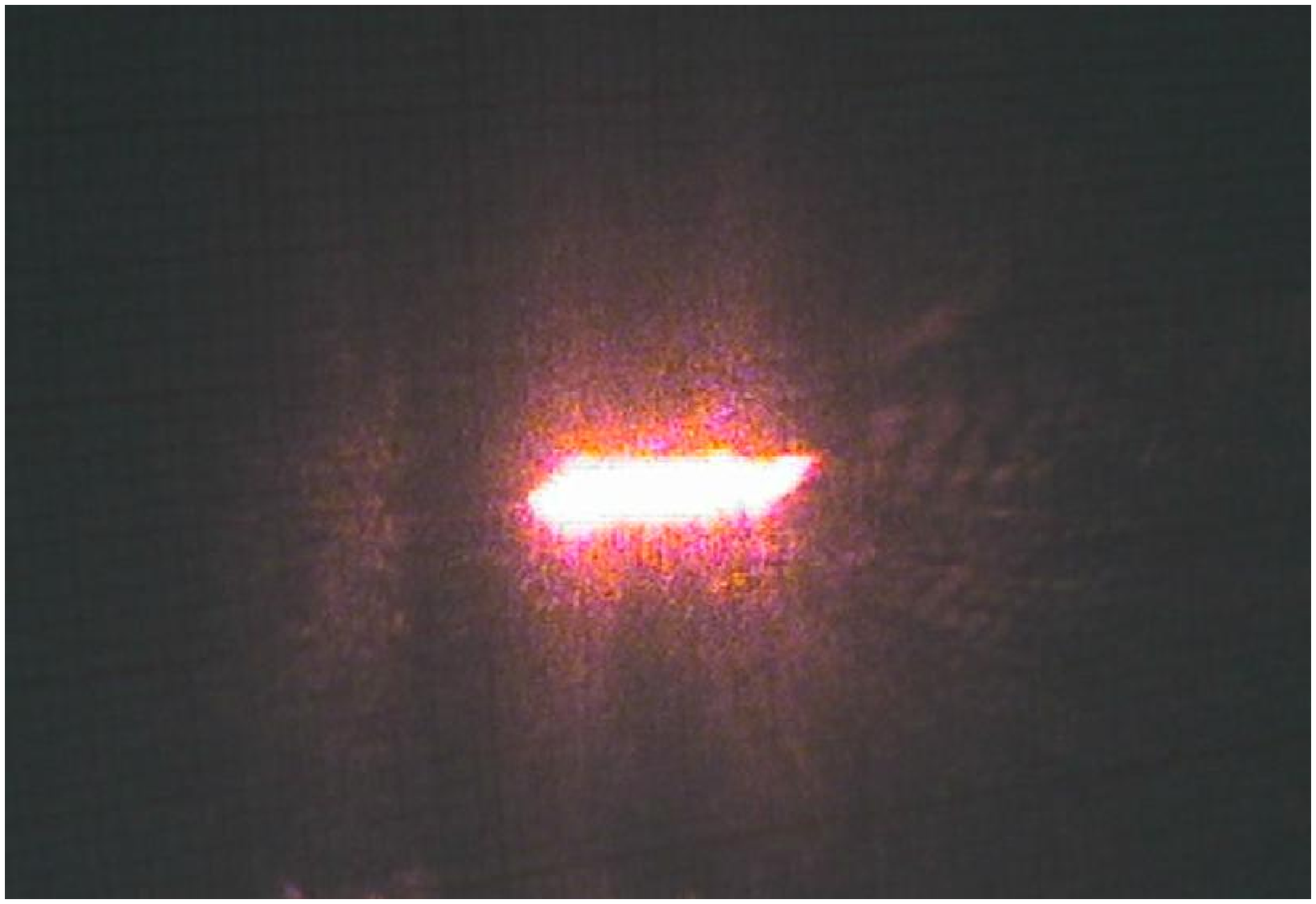} b) %e %gf12.eps
\epsfxsize=4.3cm %
\epsfysize=2.95cm %
\epsffile{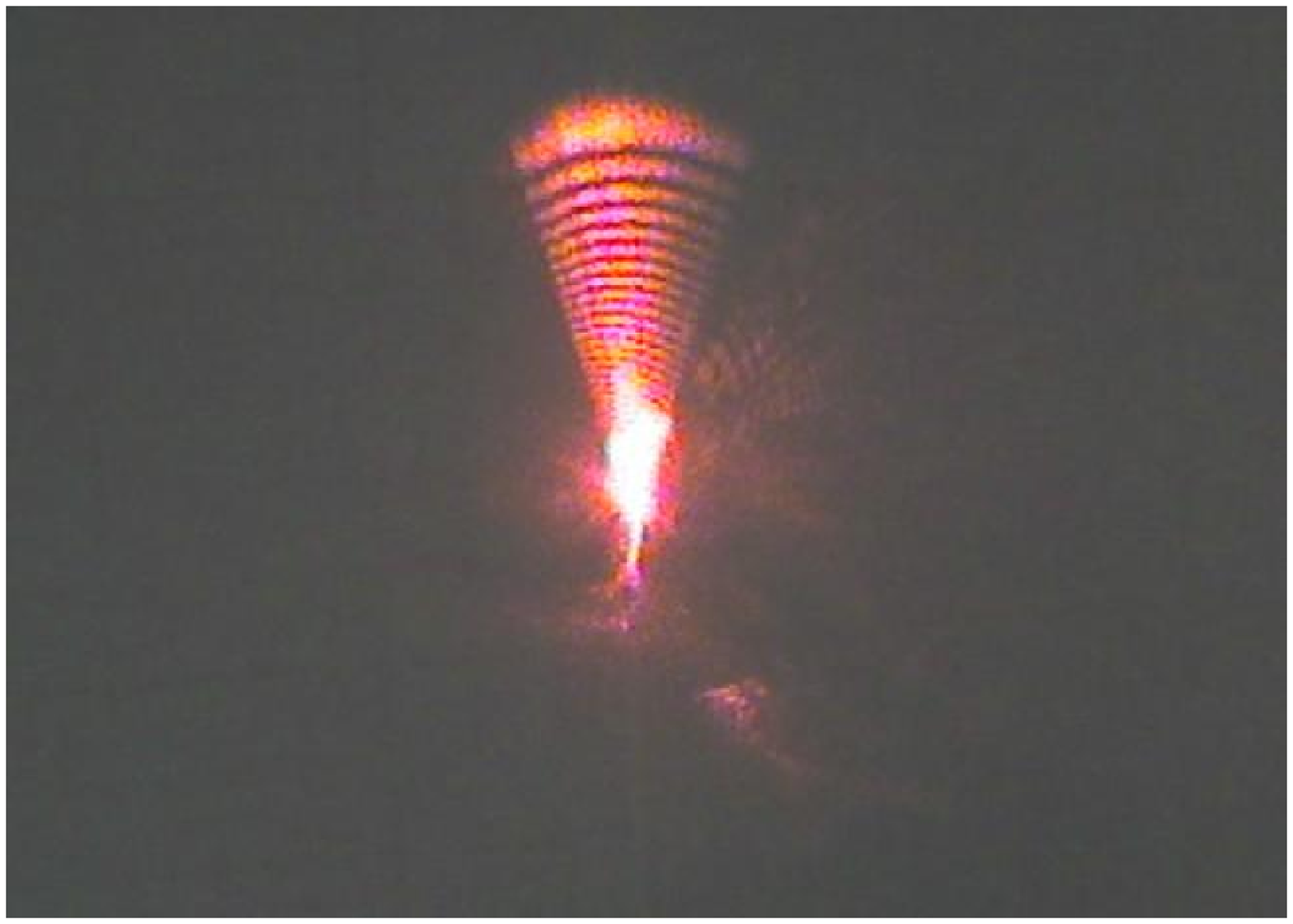} c) %f %f7.eps
\end{tabular}
\caption{a) {\em frontal view of a laser beam in the absence of any surface inside the medium}; b)-c) {\em scattering of a laser beam from a {\rm (}homogeneous and inhomogemeus, respectively{\rm )} surface of a pipe generated in glycerol.}}
\label{Fig2}
%\label{figO9}=>fig2a, fig2b
%\label{figO8}=>fig2c
\end{center}
\vspace{-0.7truecm}
\end{figure}

We have then investigated about the density and the viscosity of the liquid inside the pipe and found that their values are slightly lower than those corresponding to the bulk liquid. We have indeed observed that pipes generated in pure glycerol not in a vertical position, but anyhow inclined, slowly climb up in the bulk glycerol. On the other hand, in any working condition, the terminal velocity of spheres falling inside a pipe is always slightly larger than the velocity of spheres falling in the bulk glycerol. We have been able to set a rough upper limit of about $4\%$ on the per cent variation of such physical quantities. For example, we have obtained that:
\be %
\frac{\eta_{\rm pipe}}{\eta_{\rm bulk}} = 0.96 \pm 0.15
\label{visc} \ee %
and similarly for the density.

Then, we have observed that mechanical action can be easily performed upon the pipes generated in pure glycerol without breaking them. For example, it is noticeable that one can shift them, wound a pipe around another one, deform or even bifurcate them. It is also possible to ``close'' the pipe by ``transporting'' backward (that is, in the direction opposite to that of formation) its surface; in such a case, however, even though the pipe is re-absorbed, a perturbed crater-like zone persists (for some time) on the free surface of glycerol.

As well remarkable are the optical properties of the structure generated in the viscous liquids under considerations. The pipe, indeed, is effectively visible only along its axis (from the top), but not at its sides. When illuminated along such axis with normal (not laser) light, it becomes glossier than the bulk liquid, while a shadow is produced on a screen (with respect to the bulk liquid) when illuminated at its side. In this last case, the projection of the pipe on the screen shows a greater luminosity (with respect to the bulk) at the pipe surface, while its internal part is dark, even with respect to the bulk liquid. This can be quite easily observed when the light impinges perpendicularly both on the tube and on the pipe surface. Instead, by increasing the angle of incidence, the brighter edges increase in width ``invading'' the dark internal part, till they completely cover the internal part. In such a case, all the projection of the pipe is brighter with respect to the bulk liquid. \footnote{All such effects, including those described in the following, are however much lesser visible in castor oil.}

Even more intriguing effects are observed when a laser beam (of sufficiently high intensity) is employed to illuminate the pipe. Its light, in fact, is scattered normally to the pipe surface, the light disc on the screen becoming elliptical in its shape (see Fig.\,\ref{Fig2}). As checked, such a scattering is similar to that from a glass (or plastic) tube filled with a fluid (air, water or glycerol) and immersed into a larger tube filled with glycerol, thus denoting a well defined structure constituting the pipe. It is also remarkable that, when the pipe is not regularly (cylindrically) shaped, or some inhomogeneities are present near its surface, the laser beam scattered from such surface presents bright fringes alternated with dark ones. These are visible for any angle of incidence, even when the laser impinges on the backward lateral surface of the pipe (see Fig.\,\ref{Fig2}c).

\

\noi It is then remarkable that, just after the falling of the sphere in pure glycerol, the diameter of the pipe generated coincides with that of the sphere but, very rapidly, the pipe becomes thinner till its diameter takes a stationary value (largely lower than the diameter of the sphere). However, we have noted that such a thinning takes place by means of liquid re-climbing around the pipe towards the top of the tube, rather than by means of a narrowing/absorption process. 

Such a process, instead, takes place when the described non-stationary effect comes to an end, and quantitative observation can be quite easily carried out. Measurements of the diameter of the pipe divided by that of the falling sphere ($d_{\rm ratio} = d_{\rm pipe}/d_{\rm sphere}$) at given instants of time since pipe formation (chosen as the reference time $t=0$) have been performed as a function of the concentration of glycerol/water mixtures. Such observations have been obtained by illuminating the sample with normal (non laser) light and measuring the size of the shadow of the pipe and the sphere projected laterally on a screen. A given measurement ended when an appreciable shadow (with respect to the bulk liquid) was no more projected on the screen. Notice, however, that at this time the pipe continued to persist, as it has been observed by looking directly at the pipe from its top, see below measurements have been performed with spheres of diameter 5 mm, at a temperature of $24 \div 25 ^{\rm o}$C.)

\begin{figure}%8 NUOVA FIGURA 3
\begin{center}
\begin{tabular}{ccc}
\epsfxsize=6.6truecm %
\epsffile{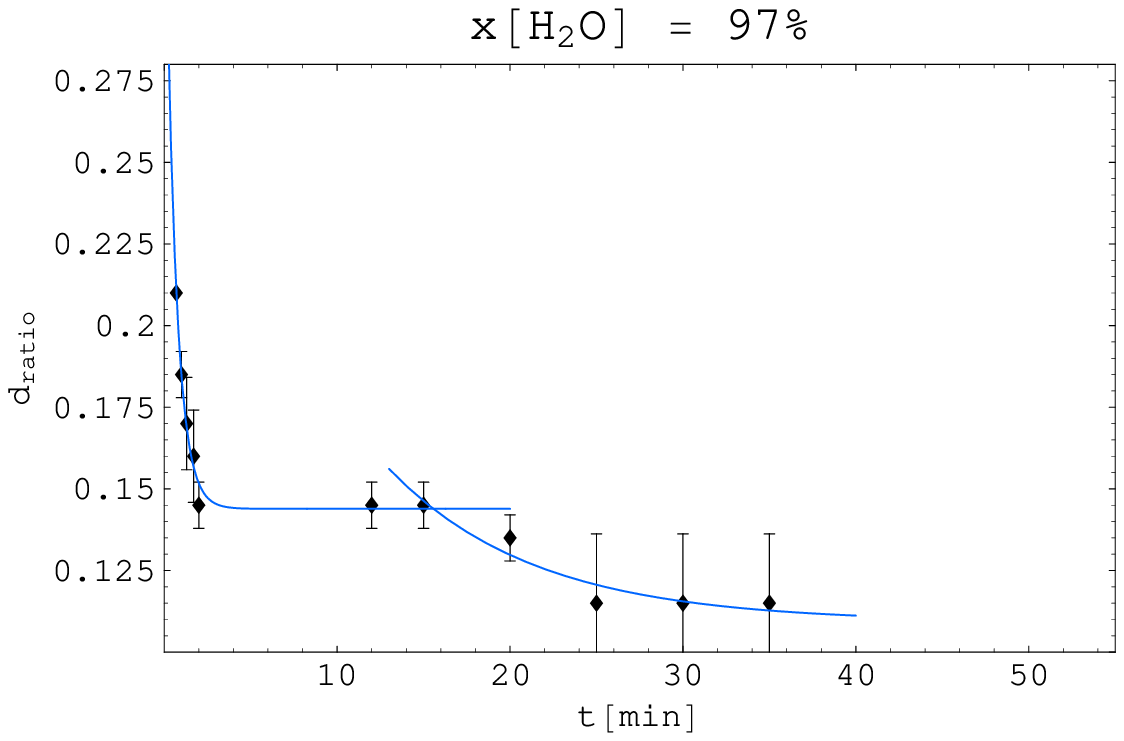} & &   %c)   d97.eps
\epsfxsize=6.6truecm %
\epsffile{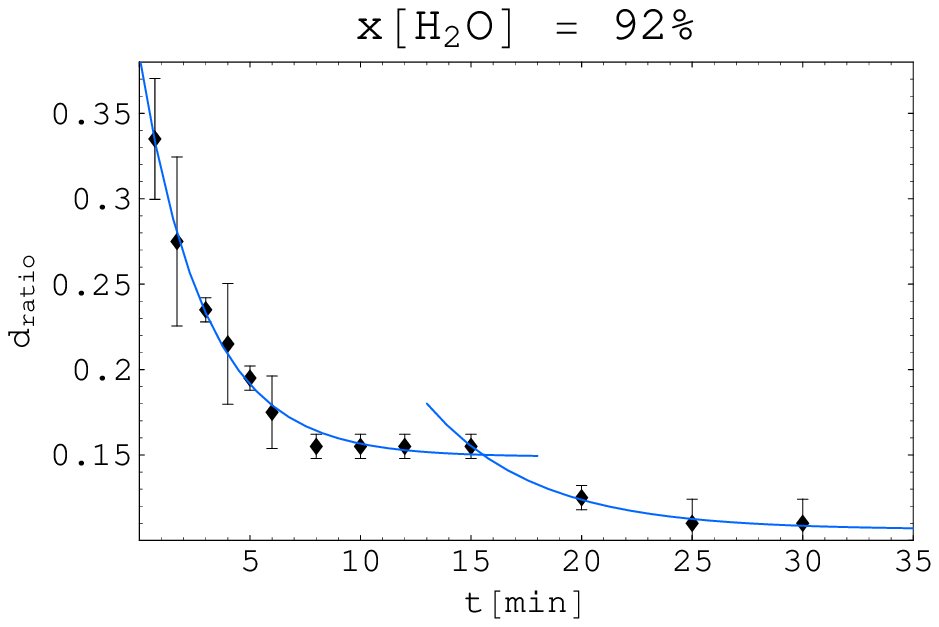} \\    %h)   d92.eps
\end{tabular}
\caption{Time evolution of the pipe radius for different water
contents of the glycerol/water mixture.}
\label{Fig3}
\vspace*{-0.8truecm}
\end{center}
\end{figure}

The time evolution of the pipe radius has always an
exponentially decreasing behavior (in given regions of time), for
any concentration of the glycerol/water mixture: 
\be \label{eqdratio} %
d_{\rm ratio} = d_0 \left( \erm^{-\frac{t - t_0}{\tau}} + 1
\right) .
\ee %
In Fig.\,\ref{Fig3} we show some experimental data together with the fitting curve from Eq.\,(\ref{eqdratio}). The experimental points refer to averages over several sets of
measurements, performed at given concentrations; the big error bars occurring in some cases are due to the (relatively) big deviation between these sets of data.

\noi Another notable result obtained during our investigations is that, as anticipated above, the dynamical evolution of the pipe generated in glycerol/water mixtures takes place in different, subsequent time phases, each one characterized by its own exponential behavior. The trigger of such phases depends on the concentration of the mixture. 

\

\noi The pipe generated in viscous liquids lasts for some time, preserving its own features, depending on the properties of the liquid employed. We define a {\it persistence time} as the time from the formation of the pipe till its visible
disappearance, that is, till the intensity of the light reflected
by the pipe and revealed from the top (parallel to the pipe axis)
reduces below a given experimental threshold. As expected, we have observed that it strongly depends on the viscosity of the liquid employed. For example, when glycerol/water and glycerol/ethanol mixtures are used, it directly depends on the concentration of glycerol: by lowering it, the persistence time gets shorter (see below). A similar behavior has been observed in castor oil as well, although the persistence time in such a case is much shorter than for glycerol.

Non-negligible errors come out when measuring the persistence time, due to the fact that, before the pipe disappears, it gets deformed, its shape becoming no more cylindrical with a straight line axis. Such a deformation apparently depends on the (horizontal and vertical) sizes of the tubes employed. This holds for pure glycerol as well as for glycerol/water and glycerol/ethanol mixtures, although the deformation in such mixtures is more accentuated (for lower glycerol concentrations).

\begin{figure}%9 NUOVA FIGURA 4	
\begin{center}
\epsfxsize=6.5cm %
\epsffile{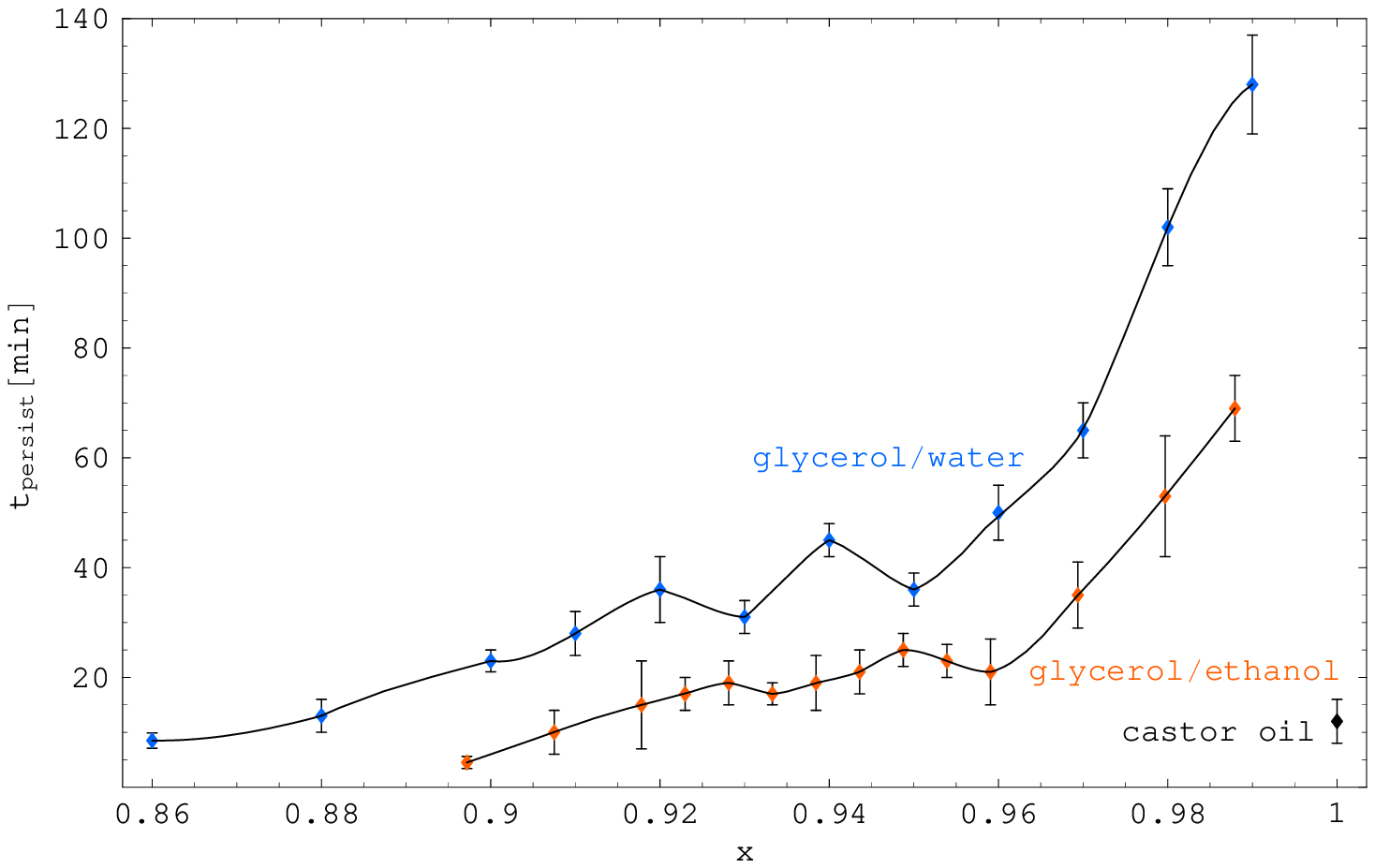} a)   %watethancast.eps
\epsfxsize=6.5cm %
\epsffile{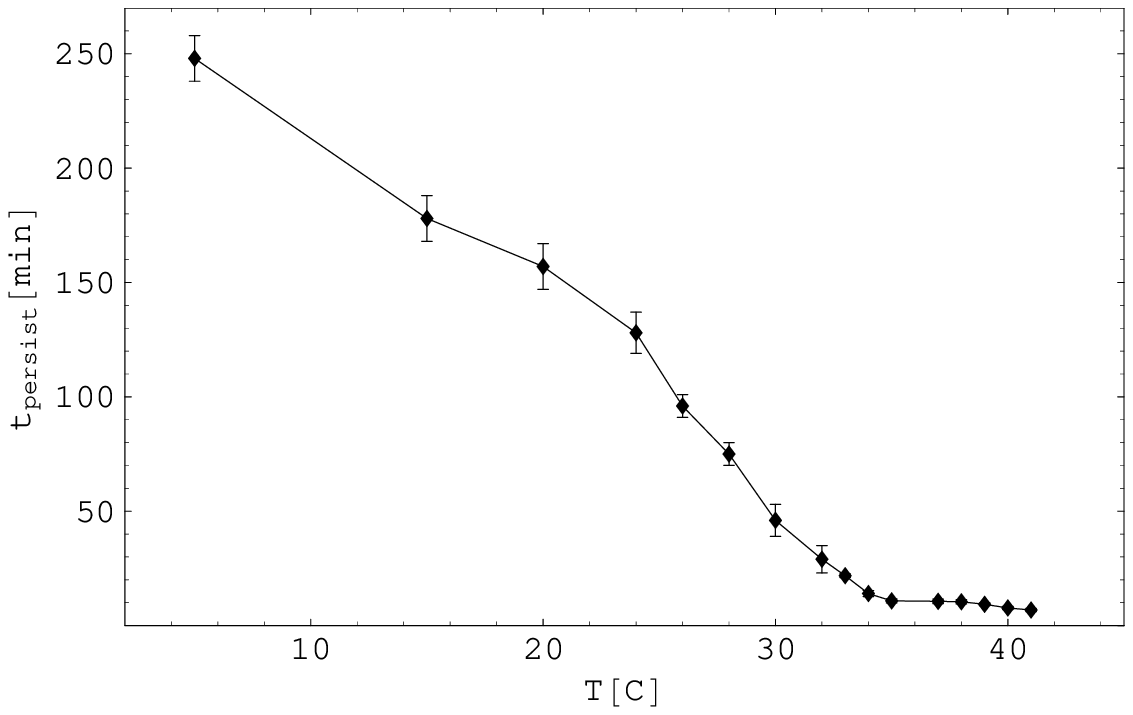} b)   %tpt.eps 
%: interpolation of the experimental points {\em (}upper plot{\em )} and data fitting with two different gaussian curves {\em (}lower plot{\em )}}; 
%\newline} \label{figwatethancast}
\caption{a) {\em persistence time of pipes formed in glycerol/water {\em (}$24 ^{\rm o}${\em C)} and glycerol/ethanol {\em (}$24.5 ^{\rm o}${\em C)} mixtures as a function of the glycerol concentration. For reference, the experimental point for the persistence time in pure castor oil 
{\em (}$23 ^{\rm o}${\em C)} is reported as well {\em (}the dependence on the concentration for castor oil mixtures is practically unobservable with the presently adopted apparatus, due to the extremely low value of the persistence time{\em )}}; \ b) {\em persistence time for pipes generated in pure glycerol as a function of the temperature}}
\label{Fig4}
\end{center}
\vspace{-0.7truecm}
\end{figure}

The persistence time broadly decreases with decreasing
glycerol concentration in glycerol mixtures reaching, however, two
relative maxima for values of the concentration approximately
given by:
\be \label{eqmaxw} %
x_{\rm glycerol} \simeq 0.92, \qquad \qquad x_{\rm glycerol}
\simeq 0.94
\ee %
for glycerol/water and:
\be \label{eqmaxe} %
x_{\rm glycerol} \simeq 0.93, \qquad \qquad x_{\rm glycerol}
\simeq 0.95
\ee %
for glycerol/ethanol mixtures, respectively. The experimental
points are reported in Fig.\,\ref{Fig4}a (we have used 3
mm spheres, at a temperature of $24 ^{\rm o}$C and $24.5 ^{\rm
o}$C for glycerol/water and glycerol/ethanol mixtures,
respectively).

In the evolution dynamics of the pipe for\-med in
glycerol/water and glycerol/ethanol mixtures, three different
contributions have been found in the $t_{\rm persist}(x_{\rm
glycerol})$ plot; they are quite well described by gaussian
fitting curves:
\be \label{eqpersx} %
t_{\rm persist} = t_a \erm^{- \frac{(x_{\rm glycerol} -
x_0)^2}{x_\sigma^2}} + t_b\,.
\ee %

\begin{figure}%11 NUOVA FIGURA 5
\begin{center}
\epsfxsize=10cm %
\epsffile{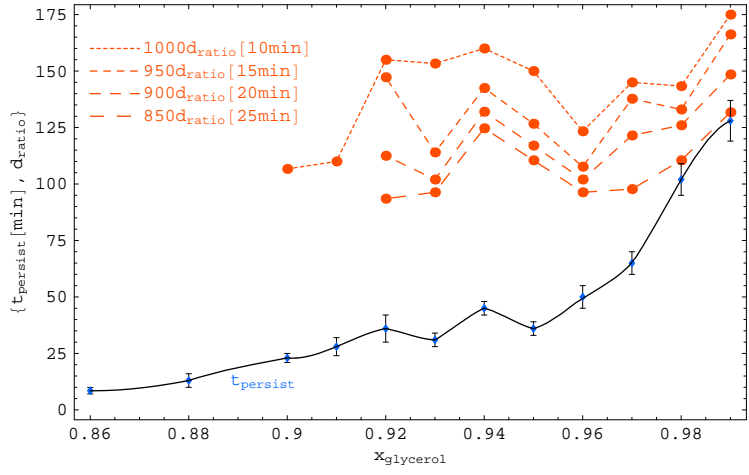}       %tpdx.eps
\caption{Comparison between the persistence
time curve (for pipes generated in glycerol/water mixtures) and
the pipe diameters curves at subsequent time instants since
formation, as a function of the glycerol concentration. The
quantities $d_{\rm ratio}$ have been multiplied by a suitable
rescaling factor (reported in the plot), in order to be easily
recognizable, while arbitrary units have been adopted on the
vertical axis.} 
\label{Fig5}
\end{center}
\vspace{-0.7truecm}
\end{figure}

Such different contributions in the evolution dynamics of the
pipe do not manifest, however, only at later times, that is in
the observed persistence time curves, but are {\it independently} confirmed by the evolution curves of the pipe diameter at different times since formation, as showed in Fig.\,\ref{Fig5}.

The pipe persistence time (in pure glycerol) decreases as well with increasing temperature, according to three different behaviors (see Fig.\,\ref{Fig4}b). In the temperature regions where enough data are available, they are quite well described by gaussian fitting curves:
\be \label{eqperst} %
t_{\rm persist} = t_c \erm^{- \frac{(T - T_0)^2}{T_\sigma^2}}\,.
\ee %
The curves $t_{\rm persist}(x_{\rm glycerol})$ and $t_{\rm persist}(T)$, obtained so far with evidently different experimental methods, can be compared when one assumes that the variations of glycerol concentration and temperature contribute essentially to change only the viscosity $\eta$ of the liquid. We have checked that the appearance of different contributions to the evolution dynamics of the pipe is effectively ruled by the viscosity parameter, though a non-negligible role is nevertheless played directly by the water (or ethanol) content of the mixture (through the concentration $x_{\rm glycerol}$) and by the thermal agitation (through the temperature $T$) for viscosity values lower than approximately 550 mPa $\!\cdot\!$\,s.
A similar behavior is shown even for pipes generated in different viscous liquids, although the phenomenon studied develops on different time scales. 

\begin{figure}%14 NUOVA FIGURA 6
\begin{center}
\epsfxsize=10cm %
\epsffile{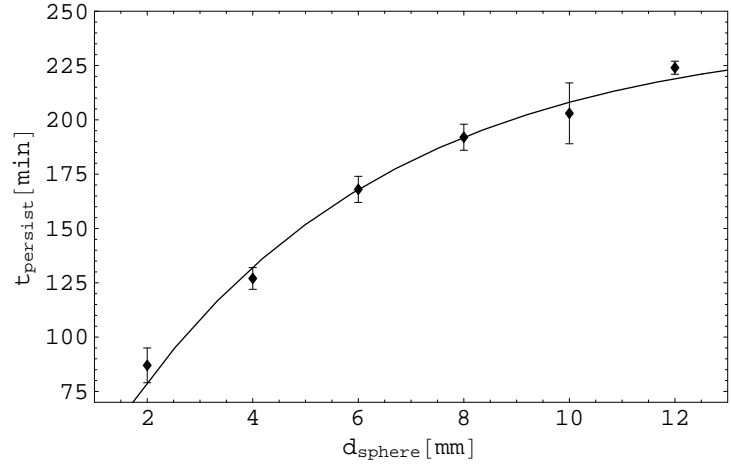}      %tpdsp.eps
\caption{Persistence time of pipes generated
in pure glycerol ($24.5 \div 25 ^{\rm o}$C) as a function of the
diameter of the generating spheres. \newline} \label{figpersdiam}
\label{Fig6}
\end{center}
\vspace{-0.7truecm}
\end{figure}

Finally, we have also measured the dependence of the persistence time of pipes formed in pure glycerol on the diameter of the falling sphere. The persistence time increases exponentially with the sphere diameter, as shown in Fig.\,\ref{Fig6}, and the fitting function appearing in this plot writes:
\be \label{eqpersd} %
t_{\rm persist} = t_d \left( 1 - \erm^{- \frac{d_{\rm sphere}}{D}}
\right),
\ee %
where $t_d = 240 \pm 10$ min and $D = 5.0 \pm 0.5$ mm.

\

\noi From the collection of qualitative observations reported above it comes out quite clearly that the falling of a heavy sphere (of
whatever diameter) induces a modification of the viscous fluid
along its path, resulting in the formation of a definite surface
bounding the pipe. The appearance of such a surface is mainly
responsible of the various mechanical and optical effects reported
above, the physical properties of the liquid used influencing the dynamical evolution of the pipe and its geometrical properties. The development of the phenomenon observed is certainly related to relaxation processes taking place in the viscous liquids employed, but these ``standard'' processes appear to be not the main source of the phenomenon itself, as shown by the quantitative results reported above.

The evolution dynamics of the pipes generated in viscous
liquids comes out to be quite similar for different substances,
but takes place on peculiar time scales and tends to be very fast
for not high values of the viscosity, so that for moderately
viscous or inviscid liquids the effect is practically
unobservable.
The phenomenon appears to be completely settled by the liquid
viscosity for \mbox{$\eta \geq 550$ mPa $\!\cdot\!$ s}, and, in this region, 
the persistence time increases monotonically with increasing
viscosity. Instead, for lower values of the viscosity, the
microscopic structure of the liquid and its thermal energy play a
non-negligible role: different contributions to the evolution
dynamics come out at any instant of time since pipe formation, and
the persistence time curves exhibit two maxima for peculiar values
of the concentration (depending on the type of mixture used). In
any of these regions, the persistence time curves are quite well
approximated by gaussian fits, irrespective of the particular
liquid mixture employed. It is particularly remarkable the fact
that the appearance of such diverse contributions is confirmed by
four independent observations (measurements of: persistence time vs
glycerol/water concentration, persistence time vs glycerol/ethanol
concentration, extinction rates at different times vs
glycerol/water concentration and persistence time vs temperature).
Finally, the extinction rates of the pipes are always exponentially decreasing with time elapsed since formation, but different time scales for the evolution of the phenomenon have been observed, the appearance of which depends on the concentration of the liquid used.

\section{Theoretical insight} \label{insight}

\noi Our experimental analysis has shown a very rich phenomenology
for the pipe effect, that cannot be simply accounted for standard
relaxation effects.

Actually, for the macroscopic size of the pipe formed \footnote{In qualitative runs we have observed that the phenomenon takes place and lasts for much time even when spheres with radius of several centimeters fall in half a meter deep containers.}, as well as for the non-trivial phenomenology observed, we can exclude an explanation for the alteration of the properties of the liquid encountered by the falling body
based on the possible dissolution of microscopic air bubbles located
in the interstices of the surface of the metal sphere (these bubbles cannot be resolved with our experimental apparatus). Possible explanations in terms of changes of the water concentration in the fluid along the path of the sphere, where this extra-water is adsorbed by the sphere on its surface when it has been sitting in air for some time, may also be considered (the glycerol is, in fact, strongly hygroscopic). However, they are ruled out as well on the same grounds, and by our direct observations that (macroscopic) water drops are rapidly absorbed by glycerol on time scales of tens of seconds, while the phenomenon observed lasts even for hours. Although spurious effects, such as the
dissolution of small air bubbles in the liquid, are sometimes
present in our observations, we have seen that the primary phenomenon of the formation of a well-defined structure is not at all related to those secondary effects.

The phenomenon of pipe formation described here appears similar to that taking place in concentrated sucrose solutions (or other similar liquids), although with a less rich phenomenology. In this case, however, the pipe is formed only when the temperature of the falling sphere is somehow (units of degree) lower or higher than the temperature of the liquid, so that the visible pipe is clearly caused by a liquid of different temperature (and therefore density and therefore refractive index). A temperature difference of about $\Delta T \sim 1^{\rm o}$C is, remarkably, also of the right order of magnitude required for explaining the apparent variation reported in Eq.\,(\ref{visc}).
The observation that no pipe is visible after the sphere stays for a long time immersed in the liquid seems to support a similar explanation also for the phenomenon observed here in glycerol, so that one is tempted to say that the sphere has a different initial temperature and heats/cools the fluid along its way down.
However, although the explanation of pipe formation in a concentrated sucrose solution in terms of a temperature difference is very reasonable (inter-molecular bonds are, indeed, easily broken just by agitation of a teaspoon, and then by a falling sphere at a different temperature), that is not the case for the effect considered here.
We have been, in fact, very careful in avoiding any heating/cooling of the falling sphere; also, we have calculated that the heat generated by the falling of a steel sphere in glycerol produces just a $\Delta T \sim 10^{-4}\,{}^{\rm o}$C, which is completely unable to explain the observed effects. Moreover, even an appreciable temperature difference between the pipe and the bulk would rapidly vanish by means of the strict thermal contact existing between them, thus not explaining at all the observed very long persistence times.

Finally, we have also excluded the possibility that the effect in glycerol is caused by electrochemical reactions with the metal: the falling sphere could leave a trail of ions which might change the density. Actually, in our experiments we have employed also spheres of different materials (or painted steel spheres), and obtained identical results.
Then, we have to search for a more complex theoretical explanation.

As a first step in the understanding of the phenomena reported
above, let us assume that the falling of an heavy sphere in
glycerol or other viscous media induces, substantially, only the
formation of a {\it separation surface} between the bulk liquid
and that in direct contact with the falling body. Possible slight
alteration of the liquid properties, as effectively observed, will
be not primarily taken into account for the moment.

Two main processes should, then, receive at least a rough
explanation: the formation of such a surface and the time
evolution of the pipe. 

\

\noi The development of a surface in polar liquids, such as water,
glycerol, etc., is characterized by a well definite orientation of
the molecular dipole moments on the said surface
\cite{physicalchemistry} \cite{bottcher} \cite{croxton}: the
degree of polarization of the surface molecules is, effectively,
greater than that of bulk ones. On the surface of the liquid, a
net polarization of the molecular dipole moments develops, with an
associated generation of a non-vanishing surface electric field
and potential. Such polarization is, evidently, due to the
presence of a polarizing electric field that for water, for
example, derives from the non-vanishing permanent quadrupole
moment of the water molecule. The polarizing electric field
generates a torque on the surface molecules that tends to
orientate the dipole moments along a given direction but, of
course, a transition zone (that can extends even over $10 \div
100$ molecular layers) exists before reaching the true bulk. More
specifically, in liquids characterized by the presence of hydrogen
bonds, domains can form where the dipoles are oriented along the
direction of maximum polarization for the given domain, while the
direction of maximum polarization of neighboring domains tend to
rotate of $180^{\rm o}$ with respect to the previous one
(similarly to the juxtaposition of north and south poles for
magnets). The presence of an electric field, then, increases the
number of ``favorably'' oriented domains at the expenses of the
those less ``favorably'' ones, until a stationary state is
reached.

Here we assume that the development of the pipe surface occurs in
a way similar to the standard one described above. Very simply,
the polarizing electric field can be thought as generated by the
``friction'' between the falling sphere and the viscid liquid,
whose value is completely non-negligible due to the high viscosity
of the liquids studied. An estimate of the order of magnitude for
the field generated in such a way can be roughly obtained as
follows.

The energy available from the falling sphere is that of the
gravitational field, accounting for \footnote{Here and in the
following, for our numerical estimates we use typical values as
used in our experimental setup; for example we take spheres of
radius $R \approx 1$ mm falling in glycerol from an height of
$\approx 10$ cm, etc.}:
\be \label{eqgrav} %
{U_G = \frac{4}{3} \, \pi \, R^3 \left( \rho_{\rm sphere} - \rho_{\rm liquid} \right) \, g \, h \simeq 2.5 \times 10^{-5} {\rm J}.}
\ee %
A part of such energy will be dissipated in the production of heat
(that is, it will increase just the thermal agitation of the
liquid) due to the slowing down action of the liquid on the
falling sphere. This energy can be calculated as the work $U_{\rm S}$ done by the Stokes force $F=6 \pi R \eta v$ during the falling:
\bea
U_{S} &=& \int_0^h F \cdot \drm z = 6 \pi R \, \eta \int_0^{t_h} v^2 \, {\rm d} t = 6 \pi R \, \eta \, v_L^2 \, \Delta t \, ,
\eea
where we have used the time law for the velocity, $v = v_L \left( 1 - {\rm e}^{-t/\vartheta} \right)$, with $v_L = 2(\rho_{\rm sphere}-\rho_{\rm liquid})g R^2/9 \eta$ the terminal velocity and $\vartheta = 2 \rho_{\rm sphere} R^2/9 \eta$ the time constant of the instantaneous velocity of the sphere. The time interval $\Delta t$ is approximately equal to the time constant $\vartheta$, so that
\be \label{eqslowing}%
U_{S} \simeq 6 \pi R \, \eta \, v_L^2 \, \vartheta \simeq 5
\times 10^{-6} {\rm J} .
\ee %

\noindent We assume that a non-vanishing energy $U_P$ available for the polarization mechanism described above exists, whose order of
magnitude is given by:
\be \label{eqpol} %
U_P \simeq U_G - U_{S} \approx 10 ^{-5} {\rm J}.
\ee %
Such energy will be used to orientate the molecular electric
dipoles, so that it will be proportional to the polarizing field
$E$:
\be \label{eqdipole} %
U_D = n \, \mu \, E,
\ee %
where $\mu$ is the electric dipole moment of the glycerol molecule
and $n$ the number of electric dipoles. 
{Eq.\,(\ref{eqdipole}) holds, in general, only for a dipoles dilute gas, so that it takes no account of correlation effects among them. However, we have checked that a more detailed treatment (see, for example, Ref.\,\cite{bottcher}) gives practically the same numerical estimate reported below, so that we will use such simple equation in our discussion.} The number of electric dipoles is {\it approximatively} given by the number of molecules intercepted by the falling sphere. The number of molecules in 1 g of glycerol is
$n_0 = N_A/M \simeq 6.5 \times 10^{21}/ {\rm g}$ ($N_A$ is the
Avogadro number, while $M$ the molecular weight of glycerol). For
simplicity we assume that the pipe formed is a cylinder, so that
$m_{\rm pipe} = \rho_{\rm liquid} \pi R^2 h \simeq 0.4 {\rm g}$
and $n = n_0 m_{\rm pipe} \simeq 2.5 \times 10^{21}$ molecules. By
equating Eq.\,(\ref{eqpol}) to Eq.\,(\ref{eqdipole}), we finally get:
\be \label{eqfield} %
E \approx 450 \, {\rm V/m},
\ee %
which is a fairly optimistic estimate, since $U_G - U_{S}$ may be
even lower by one order of magnitude (if the falling time is much
greater than the time constant $\vartheta$). Anyway, such an
estimate completely agrees with the order of magnitude of the
classic Frenkel estimate \cite{frenkel} of the surface potential
of water, so that it is quite useful, in the following, to explore
further the approximate microscopic model envisaged above.

In such a scenario, it is very likely that the liquid molecules in
the pipe and in the bulk near it experience a non-homogeneous
electric field whose maximum amplitude is reached just on the pipe
surface. It expectedly decreases away from the surface inside the
pipe, while the field is practically zero in the bulk, where the
``friction'' action by the sphere is absent.
A non-homogeneous electric field \mbox{\boldmath $ E$} exerts a translational
force \mbox{\boldmath $F$} on a given molecule,
\be \label{eqfe} %
\mbox{\boldmath $F$} = \left( \mbox{\boldmath $\mu$} \!\cdot\! \mbox{\boldmath
$\nabla$} \right) \mbox{\boldmath $E$} + \alpha \left(\mbox{\boldmath $E$} \!\cdot\!
\mbox{\boldmath $\nabla$} \right) \mbox{\boldmath $E$}
\ee %
(\mbox{\boldmath $\mu$} and $\alpha$ being the electric dipole
moment vector and the polarizability of the liquid molecule,
respectively), so that the molecules with a dipole moment pointing
along the direction of \mbox{\boldmath $ E$} will be pushed towards the regions
with a greater field intensity. In such a way, a concentration
gradient generates which induces a decrease of the density inside
the pipe, with an effect similar to that of electrostriction. From
the standard theory of such effect \cite{bottcher} we can evaluate
the variation $\Delta \rho = \rho_{\rm bulk} - \rho_{\rm pipe}$ of
the density between the pipe and the bulk, obtaining:
\be \label{eqdeltarho} %
\frac{\Delta \rho}{\rho_{\rm bulk}} \simeq \frac{E^2}{8 \pi} \,
\beta \rho_{\rm bulk} \left( \frac{\partial \varepsilon}{\partial
\rho} \right)_T \equiv \left( \frac{E}{E_0} \right)^2,
\ee %
where $\beta$ and $\varepsilon$ are the isothermal compressibility
and dielectric constant of the (bulk) liquid. As a simplifying
assumption, we take valid \cite{bottcher} the Debye theory of
electric polarization for calculating the variation of the
dielectric constant with the density,
\be \label{eqdieldens} %
\left( \frac{\partial \varepsilon}{\partial \rho} \right)_T \simeq
\frac{3}{M} \, \frac{\varepsilon -1}{\varepsilon + 2}.
\ee %
From known values we get the following estimate for the typical field $E_0$ defined in Eq.\,(\ref{eqdeltarho}):
\be \label{eqe0} %
E_0 \approx 260 \, {\rm V/m},
\ee %
which is of the same order of magnitude of the field in Eq.\,(\ref{eqfield}).

By assuming, as suggested by the preliminary experimental data
(\ref{visc}), that
\[
\frac{\Delta \rho}{\rho_{\rm bulk}} \approx \frac{\Delta
\eta}{\eta_{\rm bulk}} \simeq 4 \times 10^{-2} ,
\]
we deduce that the electric field required to generate such a
variation of the density, according to the mechanism described,
would be:
\be \label{eqee0} %
E \approx 0.2 E_0 \approx 50 \, {\rm V/m},
\ee %
an order of magnitude estimate in evident agreement with the
independent one obtained above in Eq.\,(\ref{eqfield}).

A slight increase of the liquid density near the pipe surface,
with an associated decrease inside the pipe, due to presence of
the polarizing electric field, implies a corresponding change
$\Delta \varepsilon$ of the dielectric constant of the liquid in the
pipe with respect to the bulk, as anticipated by the formulae
above. From the standard theory \cite{bottcher} we have:
\be \label{eqdeltadiel} %
\Delta\varepsilon \simeq \frac{E^2}{4 \pi} \, \beta \, \rho_{\rm
bulk}^2 \left( \frac{\partial\varepsilon}{\partial \rho} \right)_T^2
\simeq \frac{6}{M} \, \frac{\varepsilon -1}{\varepsilon + 2} \, \left(
\frac{E}{E_0} \right)^2
\ee %
(within the mentioned approximations). For $(E/E_0)^2 = \Delta
\rho / \rho_{\rm bulk} \simeq 4 \times 10^{-2}$ we get:
\be \label{eqdeldiel} %
\Delta\varepsilon \approx 2 \times 10^{-3}.
\ee %
The percentage variation of the dielectric constant between the pipe and the bulk is, thus, 
quite negligible, $\Delta\varepsilon / \varepsilon \approx 0.05\%$, so that the optical properties of the
pipe are practically equal to those of the bulk. This, however,
does not apply to the interface, i.e. on the pipe surface, where a
gathering of electric dipoles is present. In all respects, the
pipe behaves as a (almost cylindrical) dielectric shell, as indeed
observed experimentally.

\

\noi Once the surface of the pipe has been formed by the falling
of a sphere, the subsequent evolution consists just in the
thinning of the pipe itself. More precisely, according to our
observations, we can distinguish the main, most relevant, process of pipe narrowing followed by a final ``dissolution'' of the pipe, which gets deformed. In the narrowing
process the pipe keeps its structure, while it ``dissolves'' once
the pipe radius reduces to small fractions of its initial value.
Here we focus our attention on the main narrowing process, which
we will show it can be described by standard thermodynamical
procedures.

Let us assume the internal energy of the system roughly equal to the energy of
the molecular electric dipoles, given by Eq.\,(\ref{eqdipole}). Its variation
during the evolution of the pipe is described by the first law of thermodynamics:
\be
\drm U = T \, \drm S - P \, \drm V \, .
\ee
As a simplifying assumption, we suppose that no appreciable heat transfer occurs with the surroundings, and the the narrowing of the pipe proceeds adiabatically, so that $T \, \drm S = 0$. The
utilizable work $\drm W$ gained by the system, $P \, \drm V$, is essentially the work done by the liquid to induce a variation $\drm A$ of the pipe surface, which is equal to $\sigma \drm A$ ($\sigma$ being the surface tension of the viscous liquid, whose value is supposed to be equal in the bulk and in the pipe). We thus have:
\be
\mu E \, \drm n = - \sigma \, \drm A \,
\ee
For simplicity we consider the pipe as a cylinder of radius
$r$ and height $h$ much greater than the radius, so that $\drm A \simeq 2 \pi h \, \drm r$. By substituting above we obtain the equation for the variation of the number of particles $n$ during the pipe narrowing:
\be
\frac{\drm n}{\drm r} = - \frac{2 \pi \, h \, \sigma}{\mu E} \, ,
\ee
whose solution can be written simply as
\be \label{eqn0}
n(r) = n_\ast \left( 1 - \frac{r}{r_\ast} \right) .
\ee
The number of molecules in the pipe, thus, increases linearly with the decreasing of pipe radius till a maximum $n_\ast$ (ideally for $r=0$), which then corresponds to a bulk value (see below). The quantity $r_\ast$ with the dimensions of a length is given by
\be \label{eqr0}
r_\ast = \frac{n_\ast \mu E}{2 \pi \, h \, \sigma} \, ,
\ee
and arises naturally in the present model: at such a value we have $n=0$, that is to say the ideal situation where, at the onset of the falling of the sphere, no particle is in the pipe. Namely, quantity $r_\ast$ can be interpreted as the radius of the falling sphere $r_{\rm sphere}$ (on the other hand the radius of that part of the initial pipe in contact with the sphere is equal to  $r_{\rm sphere}$). The quantity $r/r_\ast$ is, thus, just what we previously denoted with $d_{\rm ratio}$. Note that Eq.\,(\ref{eqr0}), which can be rewritten as $n_\ast \mu E = \sigma A_0$ (with $A_0= 2\pi r_\ast h$), comes just from the requirement that the total energy $n_\ast \mu E$, with $n_\ast$ given by the bulk value, be transformed into the work spent to form the initial pipe surface of area $A_0$.

In the real case, the narrowing process does not effectively start at $r = r_{\rm sphere}$, neither ends at the complete vanishing of the pipe, at $r=0$. From the experimental data here collected, we have that the starting point of the process can be fixed at $r= 2 d_0 r_{\rm sphere}$ (see Eq.\,(\ref{eqdratio})), while the asymptotic end-point is at $r = d_0 r_{\rm sphere}$, so that the present model would effectively apply only between these two limits. Nevertheless, a rough estimate of the number $n_\ast$ gives
\be
n_\ast = \frac{\sigma A_0}{\mu E} \simeq 10^{22} \mbox{ molecules,}
\ee
which is a correct order of magnitude estimate (compare with the number of molecules intercepted by the falling sphere leading to the estimate in Eq.\,(\ref{eqfield})).

During the narrowing process, the density $\rho = n \, m / V$ increases correspondingly with an inverse power law behavior for $r<r_\ast$ (or, rather, for $d_0 r_{\rm sphere} <r< 2d_0 r_{\rm sphere}$):
\be
\rho (r) = \tilde{\rho} \left( \frac{1}{d_{\rm ratio}^2} -
\frac{1}{d_{\rm ratio}} \right) .
\ee
Here
\be \label{rhoref}
\tilde{\rho} = \frac{2 m \sigma}{\mu E \, r_\ast}
\ee
is the value of $\rho (r)$ at $r=2r_\ast/(1+\sqrt{5}) \simeq 0.62\,r_\ast$, and sets the reference value for the density in the pipe during its evolution. A smaller reference value would raise the density difference between the pipe and the bulk (the bulk density is, obviously, constant), thus leading to larger persistence times. From Eq.\,(\ref{rhoref}) we then deduce that, for larger falling spheres (i.e. larger $r_\ast$), a larger persistence time comes out, as effectively observed experimentally (see Fig.\,\ref{Fig6}).
Also, in glycerol/water mixtures, an increase of the water content leads to an increase of the surface tension $\sigma$ and, through
Eq.\,(\ref{rhoref}), of the value of $\tilde{\rho}$, so that the persistence time would decrease: the narrowing process is thus faster for water-rich mixtures, as it appears in Fig.\,\ref{Fig3}. However, it should be taken into account that $\tilde{\rho}$ depends on the product of the surface tension with the molecular weight (which decreases with increasing water content), so that even this last quantity enters into the game. The changing of the persistence times with the increasing of water content in glycerol/water mixture appears, then, ruled by the combined action of these two quantities, and it is quite striking that the observed maxima in Figs.\,\ref{Fig4} could be related just to such an interplay. Similar considerations (but, in this case, the effect of surface tension and molecular weight is reversed) holds for glycerol/ethanol mixtures. In any case, the dominant role appears to be played by the electric dipole energy $\mu E$ available for the process to occur, as discussed above, so that all these subdominant effects should always be related to it.

Although a more detailed microscopic interpretation of the phenomenon is called for a comprehensive explanation of the phenomenon itself, it is quite interesting that the simple theoretical considerations here developed are able to predict the main features really observed in experiments.

\section{Conclusions}

\noindent In this paper we have studied in detail the pipe effect
occurring in viscous liquids which, despite its simplicity, at the
best of our knowledge it has not been reported previously in the
literature. The effect, observed here in pure glycerol, in
glycerol/water and glycerol/ethanol mixtures, and in castor oil,
consists in the formation of a well-defined structure (a pipe)
after the passage of a heavy body in the viscous liquid, this
structure lasting for quite a long period.

A very rich phenomenology has been observed for such effect, that
cannot be trivially explained in terms of standard relaxation
and/or spurious effects like the dissolution of air bubbles in the
part of liquid encountered by the falling body.

A part of this phenomenology can be well summarized by simply
admitting that the pipe effect is primarily due to the formation
of a separation surface, upon which mechanical actions of 
different kind can be performed. The apparently non-obvious
optical effects observed are, then, just the manifestation of this
surface to light (normal or laser) impinging on it.

The time evolution of the present phenomenon is, as well, not at
all trivial, as our quantitative measurements of the extinction
rates and persistence time curves of the pipe have shown. While
the pipe gets thinner and thinner as the time goes on, it
experiences different exponentially decreasing time phases, with
different time scales, this effect being likely due to sudden
changes of the liquid pressure inside the pipe. By lowering the
viscosity of the sample liquid, the persistence time of the pipe
gets shorter, till quite a complete non-observability of the
effect for not high values for the viscosity. This transition has
been accurately studied by measuring the persistence time curves
for glycerol/water (or glycerol/ethanol) mixtures with decreasing
content of glycerol. As expected, the persistence time generally
decreases with decreasing concentration of glycerol but, quite
interestingly, two maxima have been observed for given
concentrations, this behavior being confirmed by several
independent observations.

Our study has not been limited to experimental observations, but
we have also tried to construct a theoretical model for the
formation and the evolution of the pipe, able to interpret (at
least a part of) the data found.

We have, thus, assumed that the falling body induces a
non-negligible polarizing electric field in highly viscous
liquids. It tends to orientate the molecular dipole moments of the
liquid (the samples employed are highly polar liquids, with the
presence of hydrogen bonds) in the neighborhood of the falling
body surface, thus generating a dielectric shell (the separating
liquid surface) that would be responsible of (part of) the
phenomenology observed. Indeed, order of magnitude estimates of
the polarizing electric field, for the case studied here, show
that such picture could not be far from reality, since the model
brings to definite predictions that have been effectively
observed.

The dynamical evolution of the pipe, once formed according to the
microscopic model proposed, seems, then, well described by a
thorough thermodynamical model. 
According to this, the evolution is just the macroscopic result of the balance between the work made by the surface tension term, to induce a variation of the pipe surface, and the internal energy of the molecular electric dipoles.
This model gives a likely account of some of the key features observed experimentally on the persistence times of the pipe, ruled also by the combined effect of the surface tension and the molecular weight of the liquid considered. The prediction of the changing of the persistence times with the radius of the falling sphere fits, as well, quite good with the experimental observations.

Although some understanding of the effect observed has been apparently reached, it should be regarded mainly as preliminary, since further experimental investigations are needed. Also, the interpretative models proposed for the formation and evolution of the pipe should be considered, as well, just as a starting theoretical framework, which claims for a more detailed inspection. Novel results, from both theory and experiments, are then likely to be expected.

\begin{acknowledgments}
\noi The present study was inspired by some observations from a
student of one of us (S.E.), Riccardo Frasca; our gratitude to him
and to Chiara Baldassari for essential experimental assistance is
here acknowledged. Moreover, it would not have been possible to
carry out accurately the experiments discussed in this paper
without the valuable help of G. Barigozzi, E. Biondi, E. Gatti, M. Marengo
and his group (A. Bisichini, S. dall'Olio, C. Antonini), M. Lorenzi, 
C. Pisoni, and G. Rosace.
\end{acknowledgments}

%---------------------------------------------------------------------

\end{document}